\documentclass[journal,twoside,web]{ieeecolor}
\usepackage{lcsys}
\usepackage{cite}
\usepackage{amsmath,amssymb,amsfonts}
\usepackage{algorithm}
\usepackage{algorithmic}
\usepackage{graphicx}
\usepackage{enumerate}
\usepackage{mathtools}
\newtheorem{theorem}{Theorem}[section]

\newtheorem{assumption}{Assumption}[section]

\newtheorem{lemma}[theorem]{Lemma}
\newtheorem{prop}{Proposition}[section]
\newtheorem{remark}{Remark}
\newtheorem{exmp}{Example}[section]
\newtheorem{problem}{Problem}
\usepackage{textcomp}
\def\BibTeX{{\rm B\kern-.05em{\sc i\kern-.025em b}\kern-.08em
    T\kern-.1667em\lower.7ex\hbox{E}\kern-.125emX}}
\begin{document}
\title{
Feasibility of Randomized Detector Tuning for Attack Impact Mitigation
}
\author{Sribalaji C. Anand, \IEEEmembership{Member, IEEE}, Kamil Hassan, \IEEEmembership{Student Member, IEEE}, and\\ Henrik Sandberg, \IEEEmembership{Fellow, IEEE}
\thanks{This work is supported by the Swedish Research Council under the grants 2016-00861 and 2024-00185.}
\thanks{The authors are with the School of Electrical Engineering and Computer Science, KTH Royal Institute of Technology, Sweden (email: \{srca,kamilha,hsan\}@kth.se)}}
\maketitle
\begin{abstract}
This paper considers the problem of detector tuning against false data injection attacks. In particular, we consider an adversary injecting false sensor data to maximize the state deviation of the plant, referred to as impact, whilst being stealthy. To minimize the impact of stealthy attacks, inspired by moving target defense, the operator randomly switches the detector thresholds. In this paper, we theoretically derive the sufficient (and in some cases necessary) conditions under which the impact of stealthy attacks can be made smaller with randomized switching of detector thresholds compared to static thresholds. We establish the conditions for the stateless ($\chi^2$) and the stateful (CUSUM) detectors. The results are illustrated through numerical examples. 
\end{abstract}
\begin{IEEEkeywords}
Networked control systems, Robust control, Optimization, Fault accommodation. 
\end{IEEEkeywords}
\section{Introduction}\label{sec:intro}
Owing to the increasing number of cyber-attacks on Networked Control Systems (NCS) \cite{hemsley2018history}, the security of NCS has gained increased research interest from the control community \cite{dibaji2019systems}. The recommendation to improve the security of NCS is to follow the risk management cycle, which involves three steps: risk assessment, risk mitigation, and risk monitoring \cite{ross2012guide}.

There are various methods proposed in the literature to assess the risk of attacks, such as set-based metrics \cite{murguia2020security}, performance-based metrics \cite{teixeira2021security}, simulation-based metrics \cite{giuliano2019icsrange}, etc. These metrics quantify the state degradation caused by an undetectable attacker, hereby referred to as \emph{stealthy attack impact}. Recently, \cite{urbina2016limiting} proposed a new metric that considers both the stealthy attack impact and the mean time between false alarms denoted by $\tau$ (here $\tau \geq 1$ for discrete-time systems). The main idea behind the metric in \cite{urbina2016limiting}
is briefly explained next using Fig. \ref{fig:explain}. The procedure to obtain Fig. \ref{fig:explain} is given in the appendix.
%
\setlength{\textfloatsep}{5pt}
\begin{figure}
\centering
\includegraphics[width=8.4cm]{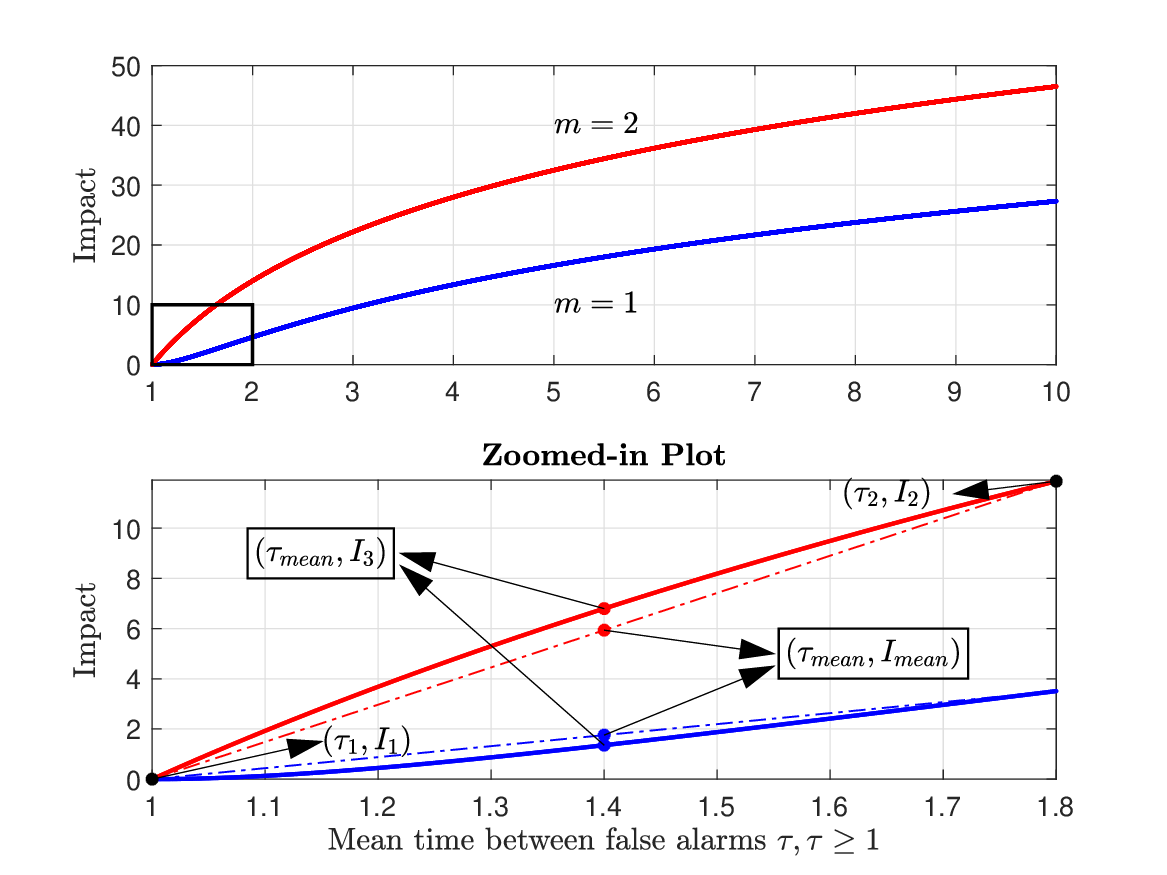}
\caption{(Top) Plot of the metric from \cite{urbina2016limiting} where the x-axis represents the mean time between false alarms $\tau$, and the y-axis represents the stealthy attack impact. Here, the plot is depicted for an NCS with a  $\chi^2$ detector and $m=1$ sensor (blue solid line) and $m=2$ sensors (red solid line). (Bottom) Zoomed-in view of the region inside the black box in the top plot. The red solid line lies above the line connecting any two points on the curve (red dotted line), representing strict concavity, whereas the blue solid line lies below the blue dotted line, representing strict convexity. When $m=2$, since the curve is concave, the impact $I_{\text{mean}}$ obtained by operating at $\tau_1$ and $\tau_2$ with probability $0.5$, is lower than the impact obtained by operating at $\tau_{\text{mean}}$. However, when $m=1$, since the curve is locally convex, the impact $I_{\text{mean}}$ is higher than the impact obtained by operating at $\tau_{\text{mean}}$.} 
\label{fig:explain}
\end{figure}

In general, the operator has two objectives. Firstly, the impact of stealthy attacks should be minimal. Secondly, when there are no attacks, $\tau$ should be large. The value of $\tau$ can be made large by setting the detector threshold $\alpha$ at a high value, which would result in a high attack impact \cite[Proposition 3]{umsonst2018game}. On the other hand, if $\alpha$ is small, the corresponding attack impact is low; however, $\tau$ would also be small as seen in Fig. \ref{fig:explain}. Thus, the operator desires a trade-off between the attack impact and $\tau$, which is the main idea behind the metric in \cite{urbina2016limiting}. This metric is an alternate to the classical Receiver operating characteristic (ROC) curve, which does not require a concrete attack hypothesis \cite{kalogiannis2022attack}.

One of the empirical observations from the metric in \cite{urbina2016limiting} is its concavity. In other words, the impact is a concave function in $\tau$. Although concavity is not theoretically proven (until now!), it is a useful result for attack mitigation. Next, we explain the usefulness of this concavity result, after which we formulate the problem studied in this paper. 
\vspace{-5pt}
\subsection{Motivation} 
Let us consider an operator with two design choices for $\tau: \tau_1, \tau_2$ where $\tau_1 < \tau_2$. Let the impact corresponding to $\tau_i$ be $I_i$ where $I_1 < I_2$. If the operator chooses $\tau_1$, the attack impact is low; however, the cost of a false alarm is high. On the other hand, if the operator chooses $\tau_2$, the cost of a false alarm is low, but the impact is high. 

Let us assume that the impact is a concave function in $\tau$. Then, at each time instant $k$, the operator can adopt a mixed strategy of choosing threshold $\tau_1$ with probability $p$, $0 < p < 1$, and $\tau_2$ with probability $1-p$. Under the mixed strategy, assuming the adversary knows the value of $\tau_1$ and $\tau_2$, the attack impact on expectation becomes $I_{\text{mean}} \triangleq pI_1+(1-p)I_2 < I_2$ and the mean time between false alarm is $\tau_{\text{mean}} \triangleq p\tau_1+ (1-p)\tau_2 < \tau_2$. If the operator chooses the threshold $\tau_{\text{mean}}$ with probability $p=1$, the corresponding impact $I_3$ satisfies $I_3 > I_{\text{mean}}$ due to strict concavity. Thus, if impact is a concave function in $\tau$, adopting a randomized strategy for choosing $\tau$ is better on average (in terms of attack impact) than a pure strategy.

Such randomized strategies are studied in the literature to mitigate attacks. For instance, the work \cite{kanellopoulos2019moving} proposes a stochastic controller switching strategy to hinder the ability to conduct a stealthy attack. The article \cite{griffioen2020moving} introduces time-varying parameters in the control system. The paper \cite{giraldo2019moving} develops an algorithm to randomly change the availability of the sensor data so that it is harder for adversaries to conduct stealthy attacks. Similar stochastic defense strategies were also adopted in \cite{umsonst2023bayesian,liu2023proactive,gallo2025switching}.

As mentioned before, adopting a randomized strategy is better if the impact is a concave function in $\tau$. However, as we observe from Fig. \ref{fig:explain}, and from \cite[Figure 13]{urbina2016limiting}, the impact is \emph{not} always a concave function in $\tau$. Thus, the main problem studied in this paper is to derive the conditions under which the attack impact is a concave function in $\tau$. By studying this problem, we present the following contributions. 
\begin{enumerate}[(1)]
\item For a $\chi^2$ detector, we provide sufficient conditions under which the stealthy attack impact is concave in $\tau$.
\item For a $\chi^2$ detector, when the number of sensors is an even integer, we provide necessary and sufficient conditions under which the impact is concave in $\tau$.
\item For a CUSUM detector with a single sensor, we derive sufficient conditions for the impact to be concave in $\tau$.
\item We extend the concavity results for the CUSUM detector to multiple sensor measurements.
\end{enumerate}

The remainder of this paper is organized as follows. We describe the NCS and formulate the problem mathematically in Section \ref{sec:problem}. The main results are presented in Section \ref{sec:results}. We depict the results through a brief numerical example. We conclude the paper in Section \ref{sec:Con}. The proofs of all the results can be found in the appendix. 

Notation: The set of real numbers is represented by $\mathbb{R}$. Given $x \in \mathbb{R}, x>0$, $\Gamma(x) = \int_{0}^{\infty} t^{x-1}e^{-t}dt$ denotes the gamma function. Given $x, y \in \mathbb{R}$ and $y >0$, the Regularised Lower Incomplete Gamma (RLIG) function is given as $P(x,y) = \frac{1}{\Gamma (y)} \int_{0}^x t^{y-1}e^{-t}dt \triangleq q$, and $P^{-1}(q;y)$ represents the inverse of the RLIG function for a given $y >0$. Given matrix $A \in \mathbb{R}^{n \times n}$, the eigenvalues of $A$ is represented by $\lambda_i(A), i=\{1,\dots,n\}$, and $\rho(A) = \max_i \vert \lambda_i \vert$ denotes the spectral radius of $A$. Given vector $x \in \mathbb{R}^n$, $x_i$ denotes the $i^{\text{th}}$ element of vector $x$, and $\|x\|_{\infty} = \max\limits_i \;\vert x_i \vert$.
\section{Problem formulation}\label{sec:problem}
In this section, we describe the closed-loop system and the adversarial policy and formulate the problem.
\subsection{System description}
Let us consider a Linear Time-Invariant (LTI) Discrete-time (DT) plant $\mathcal{P}$ described as
\begin{equation}\label{eq:plant}
\mathcal{P}: \left\{\begin{aligned}
x[k+1] &= Ax[k] + Bu[k] + \omega [k],\\
y[k] &= Cx[k] + v[k],
\end{aligned}\right.
\end{equation}
where $x \in \mathbb{R}^n$ is the plant state, $u \in \mathbb{R}^{q}$ is the control input applied to the plant, $y \in \mathbb{R}^{m}$ is the sensor measurement, $\omega \sim  \mathcal{N}(0,\Sigma_w)$ and $v \sim \mathcal{N}(0,\Sigma_v)$ represents the i.i.d. process and measurement noise respectively, $\Sigma_w$ and $\Sigma_v$ represent the positive semi-definite noise co-variance matrices, and all the other matrices are of appropriate dimension. A Kalman filter is used to estimate the states of the plant at the controller $\mathcal{C}$ 
\begin{equation}\label{eq:controller}
\mathcal{C} \left\{ \begin{aligned}
\hat{x}[k+1] &= A\hat{x}[k] + Bu[k] + L[k]r[k],\\
u[k] &= -K\hat{x}[k],\\
r[k] &= \tilde{y}[k] - C\hat{x}[k],
\end{aligned} \right.
\end{equation}
where $L$ and $K$ are the observer and controller gain with appropriate dimension, $\tilde{y}$ is the measurement signal received by the controller, and $r \in \mathbb{R}^m$ is the residue signal. Next, we establish the following.
\begin{assumption}
\begin{enumerate}
\item The tuple $(A,C)$ is observable.
\item The tuple $(A,B)$ is controllable.
\item The gain $L[\cdot]$ has reached steady state before attack.
\item $\rho(A-BK) <1$ and $\rho (A-LC) <1$.$\hfill \triangleleft$ 
\end{enumerate}
\end{assumption}

The steady state observer gain is given by $L=A\Sigma_eC^T(C\Sigma_eC^T+\Sigma_v)^{-1}$ where $\Sigma_e$ is obtained by solving the Riccati equation $\Sigma_e = A\Sigma_eA^T+ \Sigma_v -K(C\Sigma_eC^T+\Sigma_v)K^T$ \cite{milovsevic2019estimating}. Since the plant and the controller are linear, $r \sim \mathcal{N}(0,\Sigma_r)$ where $\Sigma_r = C\Sigma_eC^T+\Sigma_w$. 

The system is considered to operate nominally when $r$ is close to zero. In order to detect anomalies, we consider a detector in place. Similar to \cite{urbina2016limiting}, we consider a stateless and a stateful detector. 
\paragraph{Stateless detector}
For $\chi^2$ detector, the detection logic
\begin{equation}\label{eq:detector:chi}
\text{is}\;\; \mathcal{D}_{\chi^2}: 
z[k] \triangleq r[k]^T\Sigma_r^{-1}r[k]> \alpha(\tau) \implies \text{alarm}
\end{equation}
where $\tau \geq 1$ is the desired mean time between false alarms, and $\alpha(\tau)$ is the corresponding detector threshold. For any desired $\tau$, the threshold can be obtained using \cite[(24)]{umsonst2018game}. 
\paragraph{Stateful detector}
We consider a specific form of CUSUM detector \cite{page1954continuous} where each sensor $i, i \in  i \in \{1,2,\dots,m\}$ is employed with the detection logic
\begin{equation}\label{eq:detector:cusum}
\mathcal{D}_c: \left\{ 
\begin{aligned}
&S^{+}_i[k+1]= \max(0,S^{+}_i[k]-b+r_i[k])\\
&S^{-}_i[k+1]= \max(0,S^{-}_i[k]-b-r_i[k])\\
&S_i^{+}[0]=0, S_i^{-}[0]=0,\\
&S^{+}_{i}[k] > \alpha(\tau)\;\text{or}\;S^{-}_i[k] > \alpha(\tau) \implies \text{alarm}
\end{aligned} \right.
\end{equation}
where $b> 0$ is the bias, $\alpha(\tau)$ denotes the detection threshold common for all sensors. Such detectors are used to detect anomalies in the closed-loop system. In this paper, we consider sensor-data injection attacks described next. 
\subsection{Adversarial description}
In the closed-loop system described above, we consider a malicious adversary injecting false data. We next describe the adversarial resources, policy, and constraints. 
\paragraph{Adversarial knowledge}
The adversary knows the system matrices in \eqref{eq:plant}, \eqref{eq:controller}, the structure of the detector in \eqref{eq:detector:chi}, \eqref{eq:detector:cusum}, and the corresponding threshold. 
\paragraph{Adversarial resources}
The adversary can eavesdrop (disclosure resources) and inject data into the sensor channels (disruption resources). The attack injection is modeled as 
\begin{equation}
\tilde{y}[k] = y[k] + \tilde{a}[k],
\end{equation}
where $\tilde{a}$ is the signal injected by the adversary. However, the adversary does not have access to the actuator channels. 
\paragraph{Attack policy} We consider an attacker with a two-stage attack policy. In the \emph{first attack stage}, the adversary eavesdrops on the sensor channel to construct an accurate state estimate. In other words, during the first stage of the attack, the adversary injects no attack signal $\tilde{a}[k]=0$. Then, the closed-loop system dynamics become 
\begin{equation}\label{eq:closed:loop}
\begin{aligned}
e[k+1] &= (A-LC)e[k] + \omega[k] -Lv[k],\\
r[k] &= Ce[k] + v[k],
\end{aligned}
\end{equation}
where $e[k] = x[k] - \hat{x}[k]$. 
Once the states are estimated accurately $e[k] \approx 0$ (see \cite[Section II.B]{umsonst2018anomaly} for details), the adversary starts the second stage of the attack. In the second stage, the adversary injects an attack signal as follows
\begin{equation}\label{eq:attack:policy}
\tilde{a}[k]=-Ce[k] - v[k] + \Sigma_r^{\frac{1}{2}} {a}[k],
\end{equation}
where ${a}[k]$ is the attack signal designed by the adversary. Without loss of generality, let us assume that $\omega[k]=0, \forall k \geq 0$ (see Remark \ref{rem:sup}). Then, under the attack policy \eqref{eq:attack:policy}, the dynamics of the closed-loop system becomes 
\begin{equation}\label{eq:CL:attack}
\begin{aligned}
\begin{bmatrix}
x_a[k+1]\\
e_a[k+1]
\end{bmatrix}
&= \begin{bmatrix}
A-BK & BK\\
0 & A
\end{bmatrix}
\begin{bmatrix}
x_a[k]\\
e_a[k]
\end{bmatrix} -
\begin{bmatrix}
0\\
L
\end{bmatrix}r_a[k]\\
r_a[k] &= \Sigma_r^{\frac{1}{2}}{a}[k],
\end{aligned}
\end{equation}
where the subscript $a$ denotes the dynamics under attack.

When the matrix $A$ is not Schur, we can observe from \eqref{eq:CL:attack} that $e_a[k]$ grows unbounded for any given attack signal $a[k]$. Thus, when $A$ is unstable, the impact of sensor attacks is trivially unbounded. In order to investigate the more interesting case, we establish the following.
\begin{assumption}
$\rho(A) < 1 \hfill \triangleleft$
\end{assumption}  

\paragraph{Adversarial constraints}
We consider a stealthy adversary that does not want to raise an alarm at the detector. To this end, the adversary aims to satisfy one of the following
\begin{align}
\mathcal{D}_{\chi^2}&: r_a[k]^T \Sigma_r^{-1} r_a[k] \leq \alpha(\tau) \label{eq:stealth:chi}\\
\mathcal{D}_{c}&: S_i^{+}[k] \leq \alpha(\tau),S_i^{-}[k] \leq \alpha(\tau) & \forall \; i \in \{1,\dots,m\} \label{eq:stealth:cusum}
\end{align}
for all $k \geq 0$. Without loss of generality, we assume that the second stage of the attack \eqref{eq:attack:policy} starts at $k=0$. 
\paragraph{Adversarial objectives}
The aim of the operator is to maintain the states of the plant \eqref{eq:plant} close to zero. We then consider an adversary that aims to drive the plant states far from the origin. Before we formulate the optimization problem, we introduce some notation. 

Without loss of generality, let the adversary inject an attack for $N$ time steps (from $k=0$ till $k=N-1$). From \eqref{eq:CL:attack}, let us construct a matrix $H \in \mathbb{R}^{n\times Nm}$ such that $x_a[N]=H\mathbf{a}$ where $\mathbf{a} = \begin{bmatrix} {a}[0]^T & \dots & {a}[N-1]^T \end{bmatrix}^T$. Then, based on the detector employed, the adversary injects an attack by solving one of the following optimization problems 
\begin{equation}\label{eq:impact:chi}
\mathbb{I}_{\chi^2}(\alpha(\tau)) = \left\{
\begin{aligned} 
\max_{\mathbf{a}} & \quad \Vert H \mathbf{a}\Vert_{\infty}\\
\text{s.t.} & \quad \eqref{eq:detector:chi}, \eqref{eq:CL:attack}, \eqref{eq:stealth:chi}
\end{aligned}\right.,
\end{equation}
\begin{equation}\label{eq:impact:cusum}
\mathbb{I}_c(\alpha(\tau)) = \left\{
\begin{aligned} 
\max_{\mathbf{a}} & \quad \Vert H \mathbf{a}\Vert_{\infty}\\
\text{s.t.} & \quad \eqref{eq:detector:cusum}, \eqref{eq:CL:attack}, \eqref{eq:stealth:cusum}
\end{aligned}\right.,
\end{equation}
where $\mathbb{I}_{\chi^2}(\alpha(\tau))$ and $\mathbb{I}_c(\alpha(\tau))$ denote the impact caused against a $\chi^2$ detector and CUSUM detector, respectively. Then the main problem studied in this paper is described as follows. 
\begin{problem}
Derive the conditions under which the impact, defined in \eqref{eq:impact:chi} and \eqref{eq:impact:cusum}, is a concave function in $\tau$.$\hfill \triangleleft$
\end{problem}
\begin{remark}\label{rem:sup}
Let us denote $x_a[k] = \bar{x}_a[k]+ x_n[k]$ where $\bar{x}_a$ and $x_n$ denote the influence of attack and noise on the state, respectively. Thus, we can describe the closed-loop system in \eqref{eq:CL:attack} as the summation of two systems. Once we compute the maximum state degradation caused by attacks (say $\mathbb{I}_1$), we can compute the state degradation caused by the worst-case process noise (say $\mathbb{I}_2$). Thus, the total impact on the system would be $\mathbb{I}_1+\mathbb{I}_2$. However, since $\mathbb{I}_2$ is not a function of $\alpha(\tau)$ and/or $\mathbf{a}$, we assume that $\omega[k]=0, \forall k$. $\hfill \triangleleft$
\end{remark}
\vspace{-5pt}
\section{Results and Discussion}\label{sec:results}
In this section, we present the concavity results corresponding to a $\chi^2$ detector and CUSUM detector in Section \ref{ssec:chi} and Section \ref{ssec:cusum}, respectively. 
\vspace{-5pt}
\subsection{$\chi^2$ detector}\label{ssec:chi}
Let us consider the closed-loop system with a $\chi^2$ detector in \eqref{eq:detector:chi}. We first present the main results and then discuss the usefulness of the results presented in this section. 
\subsubsection{Main results}
Firstly, in Lemma \ref{lem:chi}, we show that the impact is a concave function in the detector threshold $\alpha(\tau)$. Secondly, in Theorem \ref{thm:main}, we provide sufficient conditions under which the impact exhibits a concave relationship in $\tau$. Finally, when ${m}$ is an even integer, we show that the impact is locally concave in $\tau$ in Theorem \ref{thm:main:iff}.
\begin{lemma}\cite[Proposition 3]{umsonst2018game}\label{lem:chi}
Let us denote the $k^{\text{th}}$ row of $H$ as $h_{k}^T$, where $h_k^T$ has $N$ partitions such that $
h_i^T\mathbf{a} = \sum_{j=0}^{N-1} h_{ij}^Ta[j]$. Then, it holds that 
\begin{equation}\label{eq:linear:chi}
\mathbb{I}_{\chi^2}(\alpha(\tau)) = \sqrt{\alpha(\tau)} f(H)
\end{equation}
where 
\begin{equation*}
f(H) \triangleq \sum_{j=0}^{N-1} \sqrt{h_{i^{*}j}^{T} h_{i^{*}j}},\;
i^* \in \underset{{i \in \{ 1, \dots, n \}}}{\arg\max}\sum_{j=0}^{N-1} \sqrt{h_{ij}^{T} h_{ij}}.\; \square
\end{equation*}
\end{lemma}

From Lemma \ref{lem:chi} and \eqref{eq:linear:chi}, we can see that the attack impact is a concave function in $\alpha(\tau)$ (since $\sqrt{\cdot}$ is a concave function in). It remains to show that that the impact is concave in $\tau$. To this end, we recall from \cite{murguia2016cusum} that for a $\chi^2$ detector, the alarm threshold $\alpha(\tau)$ is given as follows
\begin{equation}\label{eq:pf:alpha:tau}
\alpha(\tau) = 2 P^{-1}\left(1-\frac{1}{\tau};\frac{m}{2}\right).
\end{equation}
Using \eqref{eq:pf:alpha:tau}, we next present the first main result of this section. 
\begin{theorem}[Sufficient conditions]\label{thm:main}
Consider the closed-loop system \eqref{eq:CL:attack} with a $\chi^2$ detector \eqref{eq:detector:chi}, and suppose the detector threshold is given by \eqref{eq:pf:alpha:tau}. Then the impact in \eqref{eq:impact:chi} is a \emph{concave} function in $\tau$ over the domain $\tau \in [1, \bar{\tau}],$ where $\bar{\tau}$ is the largest value of $\tau$ that satisfies
\begin{equation}\label{eq:thm:con}
\frac{m}{2}-1 \geq P^{-1}\left(1-\frac{1}{\tau};\frac{m}{2}\right),
\end{equation}
where $P^{-1}(\cdot;\cdot)$ represents the inverse of the RLIG function, and $m$ is the number of sensors. $\hfill \square$
\end{theorem}

The proof of Theorem \ref{thm:main} and the other results in the sequel can be found in the appendix. 
Note that the condition in Theorem \ref{thm:main} is sufficient but not necessary. In other words, if the condition \eqref{eq:thm:con} is not satisfied, then the impact might still be concave over the domain $\tau \in [1, \bar{\tau}].$ Next, we provide necessary and sufficient conditions when $\frac{m}{2}$ is an integer.
\begin{theorem}[Necessary and sufficient conditions]\label{thm:main:iff}
Consider the closed-loop system \eqref{eq:CL:attack} with a $\chi^2$ detector \eqref{eq:detector:chi}. Let $\frac{m}{2}$ be an integer, and suppose that the detector threshold is given by \eqref{eq:pf:alpha:tau}. Then the impact in \eqref{eq:impact:chi} is a \emph{locally concave} function near $\tau$, if and only if $\tau$ satisfies 
\begin{equation}\label{m:even:con}
z! < \frac{2zz!}{\alpha(\tau)} + 2 \tau \left(\frac{\alpha(\tau)}{2}\right)^z\exp\left(-\frac{\alpha(\tau)}{2}\right),
\end{equation}
where $z=\frac{m}{2}-1$.$\hfill \square$
\end{theorem}

The conditions in Theorem \ref{thm:main:iff} are necessary and sufficient since the RLIG function admits a closed form when $\frac{m}{2}$ is an integer \cite{walter}. In other words, if $\frac{m}{2}$ be an integer, then the impact in \eqref{eq:impact:chi} is \emph{locally concave} in $\tau$ if and only if \eqref{m:even:con} holds. Next, we discuss the results presented in Theorem~\ref{thm:main:iff}. 
\subsubsection{Discussion}
\begin{figure}
\centering
\includegraphics[width=8.4cm]{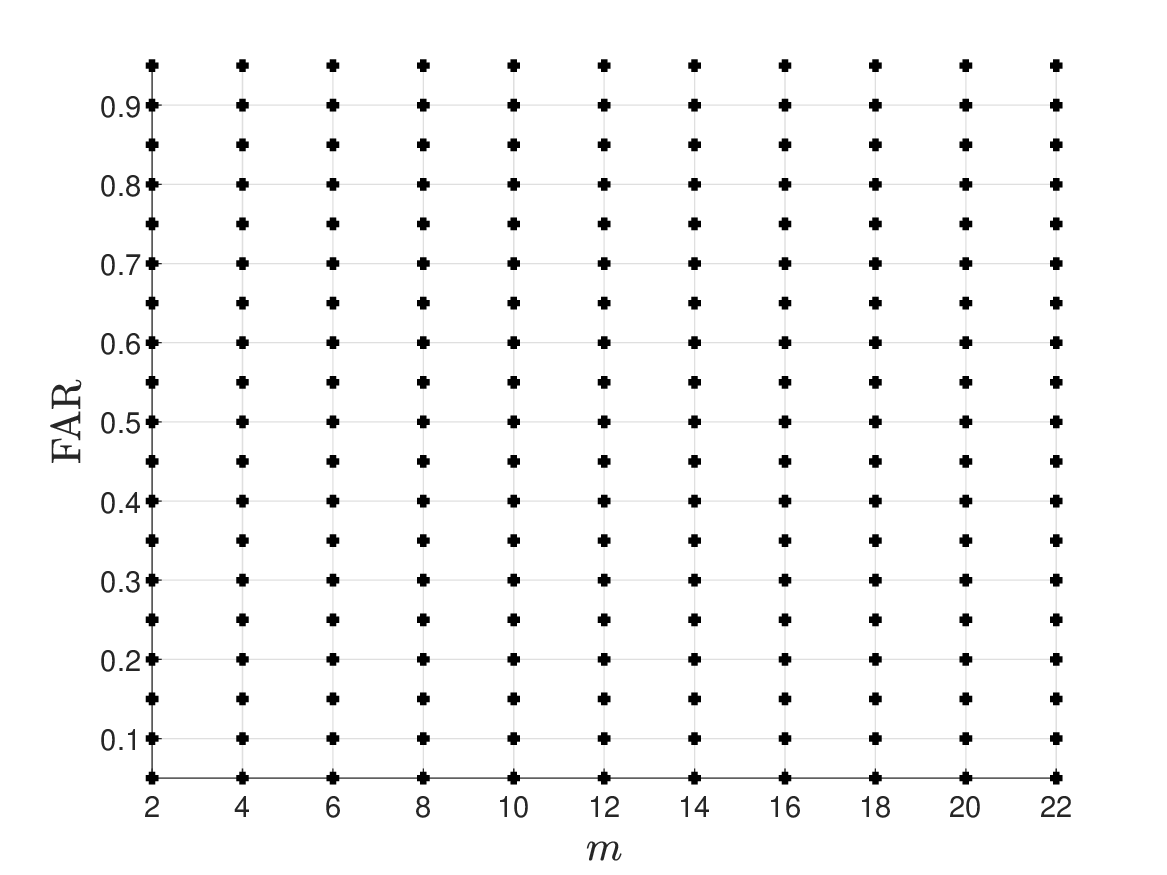}
\caption{ The black dots denotes the points where the inequality \eqref{m:even:con} is satisfied, denoting concavity.}
\label{fig:tau:bar}
\end{figure}

For various values of $\tau$, we plot the result of the inequality \eqref{m:even:con} in Fig.~\ref{fig:tau:bar} where the black dots represent the value of $m$ and $\tau$ for which \eqref{m:even:con} holds. We can see from Fig.~\ref{fig:tau:bar} that the impact is locally concave everywhere. Then, it is safe to say that the impact is usually a locally concave function in $\tau$ when $\frac{m}{2}$ is an integer, and $\alpha(\tau) >0$. Furthermore, using the Tietze-Nakajima Theorem \cite{bjorndahl2010revisiting}, we can conclude that the impact is a globally concave function in $\tau$. To briefly mention, the Tietze-Nakajima Theorem states that a subset of Euclidean space that is closed, connected, and locally convex is also convex. Next, we discuss the CUSUM detector.
%

\subsection{CUSUM detector}\label{ssec:cusum}
In this section, we consider a closed-loop system with a CUSUM detector in \eqref{eq:stealth:cusum}. We first present the main results and then discuss the usefulness of the results presented. 
\subsubsection{Main results}
In this section, we first depict that the impact is a linear function in the detector threshold $\alpha(\tau)$ and then show that the threshold is a concave function in $\tau$. 

For a system employed with a CUSUM detector \eqref{eq:detector:cusum}, the impact is given by the solution of a semi-definite program (SDP) \cite{umsonst2017security}. Since the impact does not admit a closed form expression as in \eqref{eq:linear:chi}, it is non-trivial to prove that the impact is a linear function in $\alpha(\tau)$. Thus, we consider a special case where the bias term is parameterized as $b = \delta \alpha(\tau), \delta \geq 0$ where $\delta$ is a design parameter. 

In general, $b$ is to be designed such that, as suggested by Page \cite{page1954continuous}, $S[k]$ in \eqref{eq:detector:cusum} has a negative drift when there is no attack, and a positive drift under attack. In our case, the negative drift can be achieved by tuning $\delta$. For this special case, we provide the following result. 
\begin{lemma}\label{lem:cusum}
Let $b=\delta \alpha$ where $\delta \geq 0$. Then the impact in \eqref{eq:impact:cusum} is a linear function in $\alpha$. $\hfill \square$
\end{lemma}

From Lemma \ref{lem:cusum}, we observe that the impact $\mathbb{I}_c(\alpha(\tau)) $ is a \emph{linear} function in the threshold $\alpha(\tau)$ when $b=\delta \alpha(\tau)$. To keep the presentation complete, we provide a numerical example to show the linear relationship for generic $b$.
\begin{exmp}\label{exmp:affine}
\setlength{\textfloatsep}{5pt}
\begin{figure}
\centering
\includegraphics[width=8.4cm]{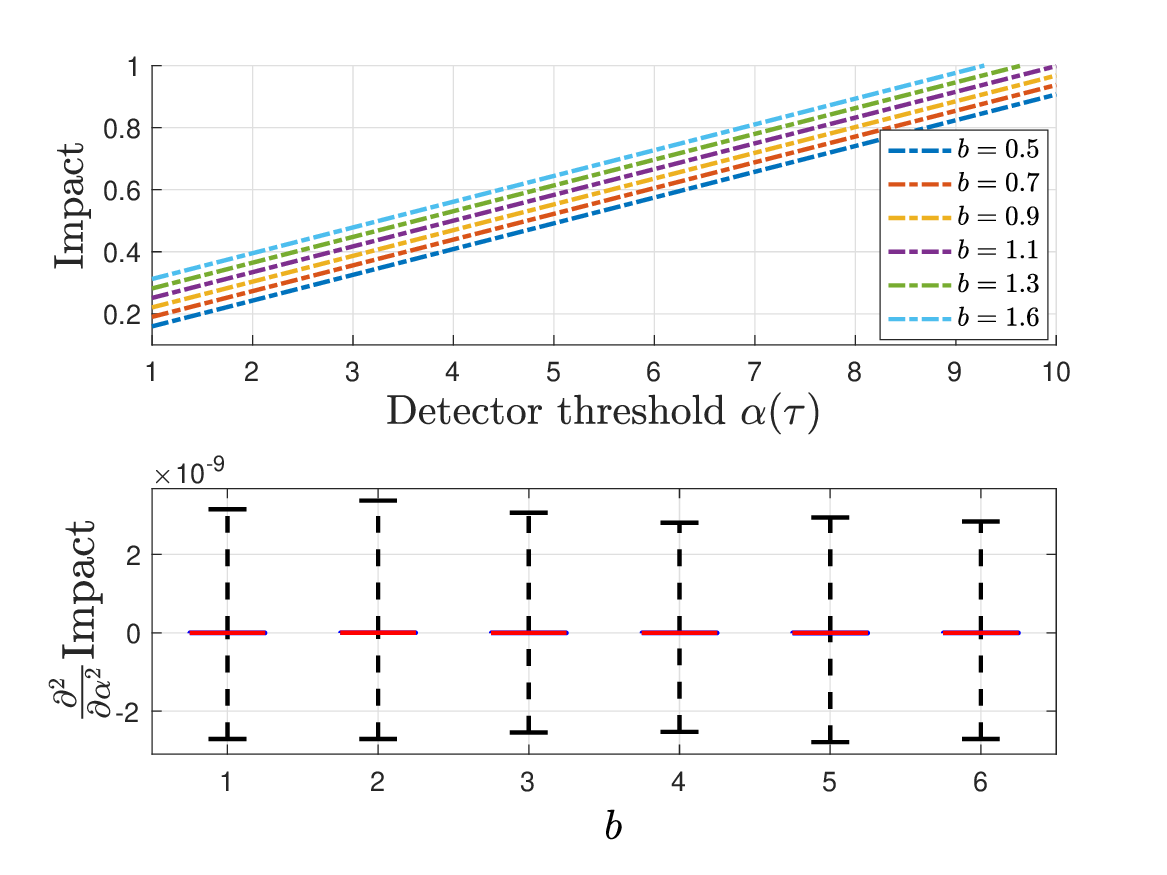}
\caption{The top figure depicts the value of impact in \eqref{eq:impact:cusum} for varying values of $\alpha(\tau)$ (increments of $0.1$) and $b$. The bottom figure depicts the box plot of the numerical second derivative of impact w.r.t. $\alpha(\tau)$.} 
\label{plot:b:neq:0}
\end{figure}
Let us consider the system \cite[(33)]{murguia2016cusum} in \eqref{eq:plant}, \eqref{eq:controller} where the system matrices are given by 
$
A=\begin{bmatrix}
0.84 & 0.23\\
-0.47 & 0.12
\end{bmatrix}$, $B=\begin{bmatrix}
0.07\\
0.23
\end{bmatrix}$, $C=\begin{bmatrix}
1 \\
0
\end{bmatrix}^T$, $
L=\begin{bmatrix}
0.25 \\
-0.18 
\end{bmatrix}$, $K= \begin{bmatrix}
1.85 \\ 0.96
\end{bmatrix}^T.$
%
%
%
%
%
%
When $N=10$, the value of the impact in \eqref{eq:impact:cusum} for varying values of $\alpha(\tau)$ and $b$ is depicted in Fig. \ref{plot:b:neq:0}. Here, the impact is obtained by solving the SDP from \cite{umsonst2017security}.
 
From Fig. \ref{plot:b:neq:0} (top), we can infer that the impact is affine in $\alpha(\tau)$ when $\alpha(\tau) \in \begin{bmatrix} 1 & 10\end{bmatrix} \triangleq \Theta$. We can also see that the numerical second derivative of the impact with respect to $\alpha(\tau)$ is close to zero ($\approx 10^{-9}$) when $\alpha(\tau) \in \Theta$. This further confirms the affine relationship between the attack impact and $\alpha(\tau)$ for any given $b$. $\hfill \triangleleft$
\end{exmp}

The above example shows that the impact can be a linear function in the detected threshold. However, proving (or disproving) the linear relationship for generic $b$ is beyond the scope of this paper. If the operator aims to choose $\alpha(\tau)$ from a discrete set, linearity can be verified numerically, as depicted in the Example \ref{exmp:affine}.

To recall, our objective is to show that $\mathbb{I}_{c}(\alpha(\tau))$ is concave in $\tau$. Under the assumption that the impact is a linear function in $\alpha(\tau)$, it is equivalent to show that $\alpha(\tau)$ is concave in $\tau$. {Here, $\tau$ can be interpreted as the Average Run Length (ARL) of a CUSUM detector in the absence of attacks before it raises a false alarm. Much effort has been made in the literature to derive a relation between ARL and $\alpha$. An exact closed-form expression is missing in the literature; however, many approximations have been proposed by Nadler and Robins \cite{nadler1971some}, Wald \cite{wald2004sequential}, Reynolds \cite{reynolds1975approximations}, and Siegmund \cite{pollak1975approximations}. A comprehensive overview of the various approaches can be found in the book \cite{basseville1993detection}. The most widely accepted and accurate approximation is the Siegmund approximation, which is also adopted in this paper.}

The Siegmund approximation can be stated as follows. Consider that there are no attacks, and let $m=1$. Then, for a CUSUM detector \eqref{eq:detector:cusum}, the mean time between false alarms ($\tau$) is given by
\begin{equation}\label{eq:ARL:whole}
\tau = \frac{\sigma_r^2}{2} \frac{\exp(2G) -1-2G}{b^2},\;G =\frac{b}{\sigma^2}\alpha(\tau) +1.166\frac{b}{\sigma_r}
\end{equation}
where $\sigma_r$ is the variance of the residue signal when there is no attack, and $b$ is the bias term. Now, we are ready to present the main result of this section.
\begin{theorem}\label{thm:cusum}
Consider the closed-loop system \eqref{eq:CL:attack} with a CUSUM detector \eqref{eq:detector:cusum}. Let $m=1$, and suppose the detector threshold is given by \eqref{eq:ARL:whole}. Then, the following statements hold.
\begin{enumerate}[(1)]
\item Let $b= \delta\alpha(\tau)$. Then the impact in \eqref{eq:impact:cusum} is a concave function in $\tau$ if $\frac{\delta\alpha}{\sigma_r} > 1$ and $\frac{\delta\alpha^2}{\sigma_r^2} > 1$.
\item Let $b$ be a positive constant, and let the impact in \eqref{eq:impact:cusum} be a linear function in $\alpha(\tau)$. Then the impact in \eqref{eq:impact:cusum} is a concave function in $\tau$. $\hfill \square$
\end{enumerate}
\end{theorem}
Next, we discuss the results presented in this section. %
\subsubsection{Discussion}
\begin{figure}
\includegraphics[width=8cm]{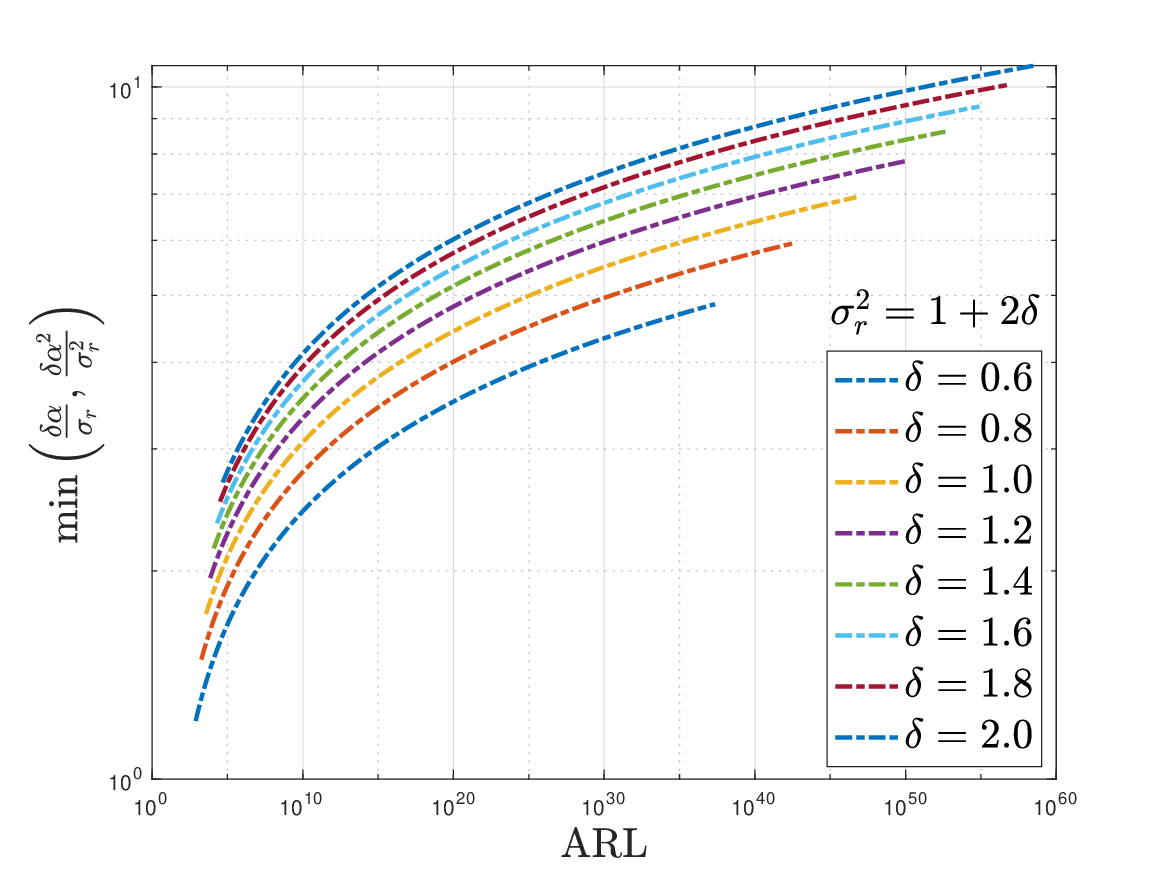}
\caption{The value of $\min\left(\frac{\delta\alpha}{\sigma_r}, \frac{\delta\alpha^2}{\sigma_r^2} > 1 \right)$ for vaing values of $\delta$. As the values are greater than $1$, concavity follows from Theorem \ref{thm:cusum} (1).} 
\label{fig:sec:der}
\end{figure}

Let us first consider the case when $b=\delta \alpha(\tau)$. Theorem \ref{thm:cusum} (1) states that the impact is a concave function when $\frac{\delta\alpha}{\sigma_r} > 1$ and $\frac{\delta\alpha^2}{\sigma_r^2} > 1$. In other words, for a given value of $\delta$ and $\sigma_r$, if $\alpha$ is chosen to be quite large, then the conditions in Theorem \ref{thm:cusum} (1) hold. We demonstrate this with a numerical results in Fig.~\ref{fig:sec:der} where we plot the ARL ($\tau$) and the minimum of the terms $\frac{\delta\alpha}{\sigma_r}$ and $\frac{\delta\alpha^2}{\sigma_r^2} $. We can see from Fig.~\ref{fig:sec:der} that for all values of $\delta$ and $\sigma_r$, the condition in Theorem \ref{thm:cusum} (1) is satisfied. Thus, for a CUSUM detector, we can safely say that the impact is a concave function in $\tau$.
However, Theorem \ref{thm:cusum} only holds for scalar measurements. Next, we extend the result to multiple measurements.\begin{prop}\label{lem:mm}
Consider the closed-loop system \eqref{eq:CL:attack} with a CUSUM detector \eqref{eq:detector:cusum}. Let $\Sigma_r = \sigma_rI_m$, $b$ be a positive constant, let the detector threshold common for all sensors is given by \eqref{eq:ARL:whole}, and let the impact in \eqref{eq:impact:cusum} be a linear function in $\alpha(\tau)$. Then the impact in \eqref{eq:impact:cusum} is a concave function in $\tau$. $\hfill \square$
\end{prop}

\section{Conclusions}\label{sec:Con}
In this paper, we studied the metric introduced in \cite{urbina2016limiting} which considered the stealthy attack impact and the mean time between false alarms $\tau$. We find non-trivial examples where the impact is not concave, meaning that static tuning is better. Under stealthy sensor attacks, we derived the sufficient (and in some cases necessary) conditions for the impact to be a concave function in $\tau$. We established the conditions for both the $\chi^2$ and the CUSUM detector. 
We also depicted our results through numerical examples. Future works include deriving concavity results for sensor attacks, and replay/routing attacks. 
\bibliographystyle{ieeetr}
\bibliography{lcsys}

\renewcommand{\thesection}{A.\arabic{section}}
\setcounter{section}{0}  
\renewcommand{\thedefinition}{A.\arabic{section}.\arabic{definition}}
\setcounter{definition}{0}
\renewcommand{\thetheorem}{A.\arabic{section}.\arabic{theorem}}
\setcounter{theorem}{0}
\renewcommand{\theassumption}{A.\arabic{section}.\arabic{assumption}}
\setcounter{assumption}{0}
\renewcommand{\theremark}{A.\arabic{section}.\arabic{remark}}
\setcounter{remark}{0}
\vspace{-10pt}
\section*{Appendix}
\section{Procedure to obtain Fig.~\ref{fig:explain}}
For any given $\tau$, the alarm threshold $\alpha$ can be obtained from \cite[(24)]{umsonst2018game}. For any given $\alpha$, we determine the impact $\left( \Vert H\textbf{a}\Vert_{\infty}^2\right)$ using \cite[Proposition 3]{umsonst2018game}. Here, the impact can be expressed as  $\alpha \times f(T)$ where $T$ is the Topelitz matrix from the attack input to the state vector, and $f(T)$ is a function independent of $\alpha$. For simplicity, we assume $ f(T)=10.1$.
\section{Appendix:Proof of Theorem \ref{thm:main}}
Before we present the proof of Theorem \ref{thm:main}, we present some preliminary results which help with the proof. 
\begin{lemma}[Proposition 1 \cite{mrvsevic2008convexity}]\label{lem:fun:inv}
If a function is monotonically increasing and convex (concave), its inverse is monotonically increasing and concave (convex). $\hfill \square$
\end{lemma}
\begin{lemma}[Section 3.2.4 \cite{boyd2004convex}]\label{lem:boyd}
Let $f(y)=h(g(y))$ and $g(y)$ be concave in $y$. Then $f$ is concave in $y$ if $h$ is concave and non-decreasing. $\hfill \square$
\end{lemma}
\begin{lemma}\label{lem:sec:der}
Let the regularized lower incomplete gamma function be represented by $P\left( \frac{\alpha}{2},\frac{m}{2} \right)$. Then, \eqref{eq:sec:der} holds. 
\begin{equation}\label{eq:sec:der}
\frac{\partial^2}{\partial \alpha^2} P\left( \frac{\alpha}{2},\frac{m}{2} \right) =\frac{\left( \frac{\alpha}{2} \right)^{\frac{m}{2}-2}}{4 \Gamma\left(\frac{m}{2}\right)} e^{-\frac{\alpha}{2}} \left( \frac{m}{2}-\frac{\alpha}{2}-1 \right).
\end{equation}
\end{lemma}
\begin{proof}[Proof of Lemma \ref{lem:sec:der}]
Recall that the regularized lower incomplete gamma function admits the form 
$
P\left( \frac{\alpha}{2},\frac{m}{2} \right)= \frac{1}{\Gamma (m/2)} \int_0^{\frac{\alpha}{2}} t^{\frac{m}{2}-1}e^{-t} dt $. Using the Leibniz integral rule, the first derivative can be obtained as 
\begin{equation}\label{eq:pf:s2}
\frac{ \partial }{ \partial \alpha} P\left( \frac{\alpha}{2},\frac{m}{2} \right) =
\frac{1}{2 \Gamma\left(\frac{m}{2}\right)}  \left( \left( \frac{\alpha}{2}\right)^{\frac{m}{2}-1}e^{-\frac{\alpha}{2}}   \right).
\end{equation}
Differentiating \eqref{eq:pf:s2} again with respect to $\alpha$ yields \eqref{eq:sec:der} which concludes the proof. 
\end{proof}
\begin{lemma}\label{lem:mon:increase}
The function $P^{-1}\left(1-\frac{1}{\tau};\frac{m}{2}\right)$ is monotonically increasing in $\tau.$ 
\end{lemma}
\begin{proof}[Proof of Lemma \ref{lem:mon:increase}]
 Consider the function $y= P(x,\frac{m}{2})$. Observe from \eqref{eq:pf:s2} that the derivative of $P(x,\frac{m}{2})$ with respect to $x$ is positive for any given $m$. Thus $P(x,\frac{m}{2})$ is monotonically increasing in $x$. From Lemma \ref{lem:fun:inv}, the function is $P^{-1}(y,\frac{m}{2})$ is also monotonically increasing in $y$. Let this be observation 1 (O1). Let $y=1-\frac{1}{\tau}$, which yields $\frac{\partial y}{\partial \tau} = \frac{2}{\tau^2} > 0$. Thus, $y$ is monotonically increasing in $\tau$. Let this be observation 2 (O2). Combining O1 and O2 and the fact that the composition of monotonically increasing functions is also monotonic, the proof concludes. 
\end{proof}
\begin{proof}[Theorem \ref{thm:main}]
It was shown in Lemma \ref{lem:chi} that the impact admits the form $\mathbb{I}_{\chi^2}(\tau)=\sqrt{\alpha(\tau)}f(H)$, where $f(H)$ is a positive constant. It follows that $$\frac{\partial \mathbb{I}_{\chi^2}(\tau)}{\partial \alpha(\tau)} = \frac{f(H)}{2}\alpha(\tau)^{-1/2}, \frac{\partial^2 \mathbb{I}_{\chi^2}(\tau)}{\partial \alpha(\tau)^2} = -\frac{f(H)}{4}\alpha(\tau)^{-3/2}.$$ Since the second derivative is negative, the impact $\mathbb{I}_{\chi^2}(\tau)$ is a concave function in $\alpha(\tau)$. Additionally, since the first derivative is positive, the impact is non-decreasing in $\alpha(\tau)$. Thus, using Lemma \ref{lem:boyd}, proving the theorem statement is equivalent to showing that $\alpha(\tau)$ in \eqref{eq:pf:alpha:tau} is concave in $\tau$. To this end, we rewrite $\alpha(\tau) = h(g((\tau))),$ where
\begin{equation}
h(g(\tau))=P^{-1}\left(g(\tau);\frac{m}{2}\right), g(\tau)=1-\frac{1}{\tau}.
\end{equation}
Since $\frac{\partial^2}{\partial \tau^2} g(\tau) = -\frac{2}{\tau^3}$, $g(\tau)$ is concave in $\tau$. From Lemma \ref{lem:boyd}, $\alpha(\tau)$ is concave in $\tau$ if $h(\cdot)$ is concave and non-decreasing. From Lemma \ref{lem:fun:inv}, the function $h(\cdot)$ is concave and non-decreasing if $h^{-1}(\cdot)$ is convex and increasing. Thus, if we show $P(\frac{\alpha}{2},\frac{m}{2})$ is convex and increasing in $\alpha$, our proof concludes. For clarity, we drop the dependence of $\alpha$ on $\tau$.

Firstly, observe that the derivative of $P(\frac{\alpha}{2};\frac{m}{2})$ with respect to $\alpha$, given in \eqref{eq:pf:s2}, is positive for any given $m$. Thus $P(\frac{\alpha}{2},\frac{m}{2})$ is monotonically increasing in $\alpha$. From \eqref{eq:sec:der}, we know that
\begin{align}\label{con:last}
\frac{\partial^2}{\partial \alpha} P\left(\frac{\alpha}{2},\frac{m}{2}\right) &= \frac{e^{-\frac{\alpha}{2}}}{4\Gamma (p/2)} \left( \frac{\alpha}{2} \right)^{\frac{m}{2}-2} \left( \frac{m}{2}-\frac{\alpha}{2}-1 \right).
\end{align}
Thus, $P(\frac{\alpha}{2},\frac{m}{2})$ is convex if $\frac{m}{2}-\frac{\alpha}{2}-1 \geq 0$. Using \eqref{eq:pf:alpha:tau}, the condition $\frac{m}{2}-\frac{\alpha}{2}-1 \geq 0$ can be reformulated as \eqref{eq:thm:con}. 

For any given $m$, let $\bar{\tau}$ represent the largest value of ${\tau}$ such that \eqref{eq:thm:con} holds. Since \eqref{eq:thm:con} holds for $\bar{\tau}$, and $P^{-1}(1-\frac{1}{\tau};\frac{m}{2})$ is a monotonically increasing function in $\tau$ (Lemma \ref{lem:mon:increase}), we can conclude that \eqref{eq:thm:con} holds for all $\tau \in [1,\bar{\tau}] \triangleq \Psi$. This concludes the proof.
\end{proof}

%
%
%
\section{Proof of Theorem \ref{thm:main:iff}}
\begin{proof}
Similar to the proof of Theorem~\ref{thm:main}, proving the theorem statement is equivalent to showing that $\alpha(\tau)$ in \eqref{eq:pf:alpha:tau} is concave in $\tau$. To this end, \eqref{eq:pf:alpha:tau} can be rewritten as
\begin{equation}\label{s1}
P\left( \frac{\alpha}{2}; \frac{m}{2}\right) = 1-\tau^{-1}.
\end{equation}
Here, the dependence of $\alpha(\tau)$ on $\tau$ is dropped for clarity. When $\frac{m}{2}$ is an integer, it follows from \cite{walter} that
\begin{equation}\label{s2}
P\left( \frac{\alpha}{2}; \frac{m}{2}\right) =1 - \exp\left(-\frac{\alpha}{2}\right)E_{z}\left[ \frac{\alpha}{2} \right],
\end{equation}
where $E_{k}[x] = \sum\limits_{i=0}^{k}\frac{x^i}{i!}$. Substituting \eqref{s2} in \eqref{s1}, we get 
\begin{equation}
\frac{1}{\tau} = \exp\left({-\frac{\alpha}{2}}\right)E_z\left[ \frac{\alpha}{2} \right].
\end{equation}
We next aim to show that $\alpha(\tau)$ in \eqref{eq:pf:alpha:tau} is concave in $\tau$ by examining its second derivative. Using implicit differentiation, we get the first derivative as 
\begin{equation}
\frac{\partial \alpha(\tau)}{\partial \tau} = 2z! \frac{\exp\left( \frac{\alpha}{2}\right)}{\tau^2 \left( \frac{\alpha}{2}\right)^z}.
\end{equation}
Let $v \triangleq \tau^2 \left( \frac{\alpha}{2}\right)^z$. Then it holds that $\frac{\partial^2 \alpha(\tau)}{\partial \tau^2} = $
\begin{equation}
\frac{2z!\exp(\alpha)}{v^2} \left[
z! - 2\tau\exp\left(-\frac{\alpha}{2}\right) \left(\frac{\alpha}{2}\right)^z - zz! \left(\frac{\alpha}{2}\right)^{-1}\right].
\end{equation}
Thus, $\frac{\partial^2 \alpha(\tau)}{\partial \tau^2} < 0$ if and only if \eqref{m:even:con} holds, which concludes the proof. 
\end{proof}
\begin{figure*}
    \centering
    \begin{equation}\label{sec:der:iff}
    \begin{aligned}
\kappa &\left(\exp(2G) \left[ \frac{6\delta\alpha^2}{\sigma_r^2} + 1.16\frac{4\delta\alpha}{\sigma_r}-3 -2\alpha^2\left(\frac{2\delta\alpha}{\sigma_r^2} + 1.16 \frac{\delta}{\sigma_r}\right)^2\right] + 3+ 2(1.16) \frac{\delta \alpha}{\sigma_r} \right)< 0\\
&\;\text{where}\; \kappa =\exp(2G)\left[ \frac{2\delta\alpha^2}{\sigma_r^2} + 1.16\frac{\delta\alpha}{\sigma_r}-1 \right] +1+1.16 \frac{\delta \alpha}{\sigma_r} 
\end{aligned}
\end{equation}
    \hrule
    \vspace{-10pt}
\end{figure*}
\section{Proof of Lemma \ref{lem:cusum}}
\begin{proof}
Consider the optimization problem \eqref{eq:impact:cusum} where $r[k]=\Sigma_r^{0.5} a[k]$ (using \eqref{eq:CL:attack}). Then, using results from \cite[Theorem 1, Proposition 3]{umsonst2017security}, \eqref{eq:impact:cusum} can be rewritten as 
\begin{equation}\label{pf:2:s0}
\begin{aligned} 
\max_{l \in \{1, \dots, n\}} \max_{\Phi} & \;\; h_l^T\mathbf{a}\\
\text{s.t.} & \;\; Q_i^{+}[0]=0, Q_i^{+}[{k+1}] \geq 0\\
&\;\; Q_i^{+}[{k+1}] \leq \alpha(\tau) \\
&\;\; Q_i^{+}[{k+1}]\geq Q_i^{+}[k] + \delta \alpha(\tau) + r[k] \\
&\;\; Q_i^{-}[0]=0, Q_i^{-}[{k+1}] \geq 0\\
&\;\; Q_i^{-}[{k+1}] \leq \alpha(\tau) \\
&\;\; Q_i^{-}[{k+1}]\geq Q_i^{-}[k] + \delta \alpha(\tau) - r[k] 
\end{aligned}
\end{equation}
where $\Phi \triangleq \{\mathbf{a}, Q_i^{+}[1], Q_i^{-}[1], \dots, Q_i^{+}[N], Q_i^{-}[N]\}$, and the constraints must hold $\forall k \in \{0, \dots, N-1\}, \forall i \in \{1, \dots, m\}$. Let us now define $\mathbf{x} = \text{vec}(\Psi) \in \mathbb{R}^{3Nm}$ and a vector $\mathbf{c}$ which is composed of $1$'s $0$'s and $\delta$, such that \eqref{pf:2:s0} becomes
\begin{equation}\label{pf:2:s1}
\max_{l \in \{1, \dots, n\}} \left\{
\max_{\mathbf{x}} \;\;  g_l^T\mathbf{x} \triangleq \begin{bmatrix}h_l^T & 0 \end{bmatrix}\mathbf{x}  \;\;\bigg \vert
\mathbf{A}\mathbf{x} \leq \alpha(\tau) \mathbf{c}
\right\},
\end{equation}
where $\mathbf{A} \in \mathbb{R}^{6Nm \times 3Nm}$. For instance when $N=m=1$, the value of $\begin{bmatrix}\mathbf{A}^T\\ \hline \mathbf{c}^T \end{bmatrix}$, and $\mathbf{x}, $ are
 \begin{equation*}
\begin{bmatrix}
-1 & 1 & -1 & 0 & 0 & 0\\
0 & 0 &0 & -1 & 1 & -1 \\
0 & 0 &  \Sigma_r^{0.5} & 0 & 0 &  -\Sigma_r^{0.5}\\
\hline
0 & 1 &  -\delta & 0 & 1 &  -\delta
\end{bmatrix}, \text{and}
\begin{bmatrix}
Q_1^{+}[1]\\
Q_1^{-}[1]\\
a[0]\\
\end{bmatrix} \text{respectively.}
\end{equation*} 

Let us now consider the inner optimization problem in \eqref{pf:2:s1}. Since it is a linear optimization problem with linear constraints, it admits a strong dual problem (See \cite[Theorem 2]{jonsson2001lecture} for details). Thus, we can rewrite \eqref{pf:2:s1} as 
\begin{equation}\label{pf:s2}
\max_{l \in \{1, \dots, n\}} \left\{
\min_{\mathbf{\lambda} \in \mathbb{R}^{6Nm}}  \;\; \alpha(\tau) \lambda^T\mathbf{c} \;\;\bigg \vert
\begin{bmatrix}g_l-\lambda^T\mathbf{A}\end{bmatrix} =0
\right\},
\end{equation}
which is equivalent to 
\begin{equation*}
\alpha(\tau) \max_{l \in \{1, \dots, n\}} \left\{
\min_{\mathbf{\lambda}}  \;\;\lambda^T\mathbf{c} \;\;\bigg \vert
\begin{bmatrix}g_l -\lambda^T\mathbf{A}\end{bmatrix} =0
\right\} = \alpha(\tau) f(H),
\end{equation*}
where the last step follows since $\alpha(\tau)$ is independent of the optimizer. This concludes the proof. 
\end{proof}
\section{Proof of Theorem \ref{thm:cusum}}
\begin{proof}
{ Proof of $(2):$ Since the impact is an affine function in $\alpha(\tau)$, proving the theorem statement is equivalent to proving that $\alpha(\tau)$ is concave in $\tau$, or showing that $\frac{\partial^2 \alpha(\tau)}{\partial \tau^2} <0$. Using implicit differentiation, it follows from \eqref{eq:ARL:whole} that 
\begin{align}
 \frac{\partial \alpha(\tau)}{\partial \tau} &= \frac{b}{\exp(2G)-1}\\
\frac{\partial^2 \alpha(\tau)}{\partial \tau^2} &= - \frac{b^3 \exp(2G)}{\sigma_r^2 (\exp(2G)-1)^3}
\end{align}
Since $b >0$, $\alpha(\tau) >0$, and $\sigma_r^2 >0$ (from the theorem statement), the term $\frac{\partial^2 \alpha(\tau)}{\partial \tau^2} $ is strictly negative, which depicts concavity. This concludes the proof of $(2)$. 

Proof of $(1):$ Since $b=\delta \alpha(\tau)$, it follows from Lemma \ref{lem:cusum} that the impact is a linear function in $\alpha(\tau)$. Thus proving the theorem statement is equivalent to proving that $\alpha(\tau)$ is concave in $\tau$, or showing that $\frac{\partial^2 \alpha(\tau)}{\partial \tau^2} <0$. Using implicit differentiation, it follows from \eqref{eq:ARL:whole} that $\dot{\alpha} = \frac{\delta^2}{\sigma_r^2}\frac{u}{v}$ where
\begin{align*}
v&=\Delta \alpha\exp(2G)-\Delta\alpha - \exp(2G) +1 +2G \\
 \quad u &= \alpha^3, \;\Delta =  \frac{2\delta \alpha}{\sigma_r^2} + 1.166 \frac{\delta}{\sigma_r}
\end{align*}
Here, we drop the argument $\tau$ from $\alpha(\tau)$ for clarity of presentation. Similarly, we derive the second derivative as 
\begin{equation} \label{pf:0}
\ddot{\alpha} = \frac{\delta^2}{\sigma_r^2}\frac{v\frac{\partial u}{\partial \tau} - u \frac{\partial v}{\partial \tau}}{v^2} \leq 0 \iff v\frac{\partial u}{\partial \tau} - u \frac{\partial v}{\partial \tau} \leq 0
\end{equation}
From the definitions of $u$ and $v$, \eqref{pf:0} can be rewritten as \eqref{sec:der:iff}. Since it holds that $\frac{\delta \alpha}{\sigma_r} >1 $ and $\frac{\delta \alpha^2}{\sigma_r^2} >1 $, we can easily show that \eqref{sec:der:iff} also holds. This concludes the proof.
}
\end{proof}
\section{Proof of Lemma \ref{lem:mm}}
\begin{proof} 
From the theorem statement, the impact is a linear function in $\alpha(\tau)$. Here, since $\alpha(\tau)$ is common to all sensors, $\Theta \in \mathbb{R}$, proving the theorem statement is equivalent to proving that $\alpha(\tau)$ is concave in $\tau$. For each sensor, if $\tau$ is desired, the threshold should be set such that \eqref{eq:ARL:whole} holds. Then, the proof follows from the proof of (1) in Theorem \ref{thm:cusum}. 
\end{proof}
\end{document}